\documentclass{mn2e}
\usepackage{epsfig,natbib2,natbibmnfix}
\usepackage{amssymb}
\usepackage{multirow}
\usepackage{multicol}
\usepackage[varg]{txfonts}
\usepackage{color}
\usepackage{url}

\newcommand{\eref}[1]{equation~\ref{#1}}
\newcommand{\cref}[1]{Chapter~\ref{#1}}
\newcommand{\sref}[1]{Section~\ref{#1}}

\begin{document}
\title[Retrograde Inflow and Merging SMBH]{Retrograde Accretion and
  Merging Supermassive Black Holes} \author[C.J.~Nixon, P.J.~Cossins,
  A.R.~King \& J.E.~Pringle] {
\parbox{5in}{C.J. Nixon$^{1\star}$, P.J. Cossins$^1$, A.R. King$^1$ and
   J.E. Pringle$^{1,2}$}
\vspace{0.1in}
  \\ $^1$ Department of Physics \& Astronomy, University of Leicester,
  Leicester LE1 7RH UK
  \\ $^2$ Institute of Astronomy, University of Cambridge, Madingley Road,
  Cambridge CB3 0HA
}

\maketitle

\begin{abstract}

We investigate whether a circumbinary gas disc can coalesce a
supermassive black hole binary system in the centre of a galaxy. This
is known to be problematic for a prograde disc. We show that in
contrast, interaction with a {\it retrograde} circumbinary disc is
considerably more effective in shrinking the binary because there are
no orbital resonances. The binary directly absorbs negative angular
momentum from the circumbinary disc by capturing gas into a disc
around the secondary black hole, or discs around both holes if the
binary mass ratio is close to unity. In many cases the binary orbit
becomes eccentric, shortening the pericentre distance as the
eccentricity grows. In all cases the binary coalesces once it has
absorbed the angular momentum of a gas mass comparable to that of the
secondary black hole.  Importantly, this conclusion is unaffected even
if the gas inflow rate through the disc is formally super--Eddington
for either hole. The coalescence timescale is therefore always
$\sim M_2/\dot M$, where $M_2$ is the secondary black hole mass and
$\dot M$ the inflow rate through the circumbinary disc.
\end{abstract}


\footnotetext[1]{E-mail: chris.nixon@astro.le.ac.uk}

\section{Introduction}
\label{intro}

Astronomers now generally agree that the centre of every reasonably
large galaxy contains a supermassive black hole (SMBH). Moreover the
mass of this hole correlates (at least at low redshift) with
properties of the host galaxy
(\citealt{FandM2000}; \citealt{Gebhardtetal2000}; \citealt{HandR2004}). In the hierarchical
picture of structure growth, small galaxies merge to produce large
ones, promoting accretion on to their central SMBHs, and apparently
causing these holes to coalesce.

The favoured mechanism for driving the holes closer is dynamical
friction. However it is unclear that this can bring them close enough
for gravitational wave losses to complete the coalescence, since the
frictional process itself scatters away the stars causing it, and
refilling of the loss cone is apparently too slow. This is often
called the `final parsec problem', as dynamical friction typically
stalls at such separations between the holes \citep{MandM2003}.

A possible way of overcoming this problem is interaction with gas
orbiting in a disc just outside the SMBH binary
(\citealt{ArmNar2005}; \citealt{MandM2008}; \citealt{Lodatoetal2009}; \citealt{Cuadraetal2009}). There has also been
much discussion of cases where an SMBH binary is embedded in a disc
(\citealt{Escala2005}; \citealt{Dottietal2007},2009). It is implicitly assumed
that 
dissipative torques make the disc coplanar with the binary. Studies of
the circumbinary disc problem have so far considered prograde discs,
i.e. those rotating in the same sense as the binary. Then tidal
interaction with the binary turns the disc into a decretion disc,
which transports angular momentum outward, but with little inward mass
transport \citep{Lodatoetal2009}. If the disc mass is large enough to
carry away the binary angular momentum, a decretion disc is vulnerable
to the self--gravitational instability \citep{Lodatoetal2009}. This
can rob the disc of the gas it needs to drive further angular momentum
loss, halting the binary shrinkage.

This makes it doubtful that a prograde disc can ever in practice
shrink the binary separation from $a\sim 1$~pc to the point ($a \sim
10^{-2}$~pc) where gravitational wave losses can drive it to
coalescence. However the separation of the SMBH binary is much smaller
than the interacting galaxies themselves so it is highly unlikely that
the central gas flows are always prograde. These flows also receive
randomly--directed injections of energy and momentum from star
formation and supernovae, suggesting that retrograde flows are as
likely as prograde. This kind of chaotic accretion gives a plausible
picture of the mass and spin evolution of a central accreting black
hole (\citealt{KandP2006};\citealt{Kingetal2008};\citealt{Hobbs}) (but see
\citealt{Bertvol}; \citealt{Mayer2007}
for an alternative picture). It also implies 
that if nothing else manages to drive an SMBH binary to coalescence,
it is highly likely that at some point there will be a retrograde
coplanar disc surrounding the binary. This situation would arise if
for example an earlier episode of prograde gas accretion failed to
coalesce the binary, and was followed by a retrograde accretion
event. As we will see, retrograde discs behave quite differently from
prograde ones, and may offer a solution to the final parsec problem.

\section{Prograde versus Retrograde}

We start by contrasting the main qualitative features of the prograde
and retrograde cases.  These stem from the physics of the interaction
between the binary and the disc, where dissipative torques try to
share the angular momenta. If the disc is prograde, this interaction
shrinks the binary but moves the inner edge of the disc outwards,
reducing the torque shrinking the binary. If instead the disc is
retrograde, the effect is to shrink the binary, but also to move the
inner edge of the disc {\it inwards} (see Appendix~\ref{appA}).

A prograde disc becomes a decretion disc, transporting angular
momentum outwards with little net mass transport
\citep{Lodatoetal2009}. A retrograde disc instead remains an accretion
disc, transporting angular momentum outwards and mass inwards. As we
shall see, the long--term evolution of the disc--binary system is
radically different in the two cases.

In particular, the disc--binary torque is quite different. For a
prograde disc the tidal interaction occurs mainly through resonances.
These occur when
\begin{equation}
\Omega^2 = m^2 (\Omega - \omega)^2,
\label{res}
\end{equation}
where $\omega$ is the binary orbital frequency, $\Omega(R)$ is the
disc angular velocity and $m = 1,\,2,\,...$ is the wave mode number
(for example see the analysis in \citealt{PandP1977}). So there are 
resonances at radii where
\begin{equation}
\Omega(R) = {m\omega\over m \pm 1}, 
\label{resradii}
\end{equation}
where $\Omega$ and $\omega$ have the same sign. Resonances outside the
binary orbit (i.e. with $\Omega < \omega$) correspond to the positive
sign in the denominator of equation~\ref{resradii} and so appear at radii where
\begin{equation}
{\Omega(R)\over\omega} = {1\over 2},\, {2\over 3},\, {3\over 4},\, ...
\end{equation}
The dominant interaction then involves the 2:1 (more strictly 1:2)
resonance.

By contrast, in a retrograde disc $\Omega$ and $\omega$ have opposite
signs, and equation~\ref{res} requires $|\Omega| > |\omega|$, so there are no
resonances in a circumbinary retrograde disc. The disc--binary
interaction is direct, as the inner edge of the disc starts to impinge
on the secondary black hole. A retrograde circumbinary disc remains an
accretion disc whose material is gravitationally captured by the
binary, directly reducing its angular momentum. This is inherently
more promising for shrinking the binary towards coalescence than the
prograde case, where the binary dams up the disc. 

It is important to understand that `capture' simply means that the gas
orbits a particular hole, and so has added its (negative) angular
momentum to the binary orbit. It does {\it not} imply that the
relevant hole must actually accrete this gas (although it may). Once
captured the gas is bound to the hole and thus may be treated as a
single body. Some or all of this captured gas can be expelled, for
example by radiation pressure. Provided that this process is isotropic
in the frame of the hole this does not change its orbital angular
momentum, and so has very little effect on the orbital dynamics and
eventual coalescence (cf equation~\ref{oneminuse} below).

\section{Where does the mass go?}
\label{retrobehaviour}

The effect of a retrograde circumbinary disc differs in detail
depending on how the captured mass is distributed between the two
black holes. Accordingly we look at the reaction of test particles to
the binary. To make things simple we first consider a circular binary with a
low mass ratio, i.e.  $M_{2}/M_{1} 
\lesssim 0.1$. Then to first order we can treat the primary as fixed
and the secondary as following a circular orbit of radius $a$ around
it with velocity $V = (GM_1/a)^{1/2}$.  The circumbinary disc gas is
on circular orbits with velocity $\simeq (GM_1/R)^{1/2}$ at each
radius $R$. We assume that the disc creeps slowly inwards from a large
radius by the usual viscous evolution. As it is counter--rotating we
can ignore all resonant effects and assume everything is ballistic
until orbits begin to cross and fluid effects appear. We define the
effective radii $R_1, R_2$ of the two holes as the radii within which
gas particles are captured by each hole, e.g. by forming a
disc around one or other of them, so that this captured gas has the
same net specific angular momentum as the relevant hole. Evidently
$R_1, R_2$ cannot in practice be larger than the individual Roche
lobes for each hole.

For the disc particles orbiting closest to $M_{2}$ the interaction
with $M_{2}$ is initially hyperbolic and so we can use the impulse
approximation. Here the relative velocity is approximately
$2V$. If the disc edge is at a radius $R = a + b$, where $b
\ll a$, the impulse approximation shows that the disc
particle acquires an inward radial velocity
\begin{equation}
   U_{R} = \left(\frac{GM_{2}}{b^{2}} \right) \times \frac{2b}{2V} =
   \frac{GM_{2}}{bV}.
\label{Ur}
\end{equation}
The inner edge of the circumbinary disc is not significantly perturbed
if its distance $b$ from the secondary's orbit is large enough that
$U_R \lesssim V$, i.e. $V^2 \gtrsim GM_2/ b$, or equivalently $b
\gtrsim (M_2/M_1)a$. From this we conclude that the secondary cleanly pulls
the gas from the 
unperturbed inner edge of the disc at $b$ provided that $R_2 \gtrsim
b$, i.e.
\begin{equation}
{R_2\over a} \gtrsim {M_2\over M_1}.
\label{rcrit}
\end{equation}
Comparing this with the Roche lobe constraint 
\begin{equation}
{R_2\over a} \lesssim 0.4f\left({M_2\over M_1}\right)^{1/3},
\label{roche}
\end{equation}
where $f \lesssim 1$ is a dimensionless factor, we see that 
the secondary captures almost all the gas
for mass ratios 
\begin{equation}
q = {M_2\over M_1} \lesssim q_{\rm crit} = 0.25f^{3/2} \simeq 0.25.
\label{qcrit}
\end{equation}
Equivalently we can define a Safronov number \citep{Safronovbook} as used in
discussing accretion of planetesimals from a 
protoplanetary disc:
\begin{equation}
\Theta = {v_{\rm esc}^2\over 2 v_{\rm orb}^2} \simeq {GM_2\over
  R_2}{a\over GM_1} = {M_2\over M_1}{a\over R_2}
\label{saf}
\end{equation}
where $v_{\rm esc}, v_{\rm orb}$ are the escape velocity from the
secondary's effective radius, and its orbital velocity respectively.
This measures how much the gas is gravitationally perturbed before
being captured. Small $\Theta$ implies little perturbation.  We see
that the criterion for secondary capture is just $\Theta \lesssim 1$,
that is, the gas is cleanly captured without significant perturbation.

For larger mass ratios the flow becomes more complex and it is likely
that some gas falls towards the binary centre of mass, producing some
form of primary capture. In numerical simulations (\sref{Numerics}) we
will find that even in this case most of the gas is captured by the
secondary. So in general the secondary captures most of the mass, and
only in a major merger with $q > q_{\rm crit}$ does the primary
capture a non--negligible amount from a retrograde circumbinary disc.

\section{Eccentricity growth}
\label{sec4}

A retrograde circumbinary disc can decrease both the energy and
angular momentum of the SMBH binary, and so change its eccentricity. A
simple argument shows how this happens. We consider a slightly
eccentric binary orbit (the orbit is never exactly circular if it is
shrinking) and again for simplicity assume that the mass ratio is
sufficiently extreme that we can regard the primary black hole as
effectively fixed, and only the secondary as interacting with the
disc. None of these restrictions will affect our conclusions (see
Appendix~\ref{appA} for a fuller discussion).

Momentum conservation shows that capture of disc gas always
reduces the secondary's orbital velocity (see
equation~\ref{mom}). This holds both for a direct collision, or (more
commonly) if the secondary captures gas into a bound disc around
itself. The secondary's mass cannot decrease in these
interactions. Its specific orbital energy therefore always drops, so
that the binary semimajor axis $a$ decreases. But for mass capture
into a symmetrical disc around the secondary near apocentre, the new
orbit must retain the same apocentre as the old one (it must pass
through this point, and the radial velocity remains zero there). Given
the decrease of the semimajor axis, this means that capture near
apocentre tends to increase the orbital eccentricity $e$ (since $a[1 +
  e]$ remains constant). The eccentricity is evidently 
\begin{equation}
e \sim {\Delta M\over M_2} 
\label{ecc}
\end{equation}
where $\Delta M$ is the amount of mass the secondary has captured.
Exactly the same reasoning shows that for mass capture near
pericentre, the quantity $a(1 - e)$ has to stay constant despite a
further decrease in $a$ -- in other words, the eccentricity must {\it
  decrease} here by about the same amount ($\Delta M/M_2$) it increased
at apocentre.

The secondary obviously captures at all points in between apo-- and
pericentre, but the effects are opposed for significant times. If the
eccentricity is initially small these times are nearly equal, and $e$ 
stays small as the orbit shrinks, provided that $\Delta M/M_2$
is below a certain value we will derive shortly. 

If on the other hand the orbit is initially quite eccentric, or the
mass grows significantly in one orbit, the pericentre may be too small
to allow disc interaction. Then $e$ grows with $a(1 + e) = a_0 \simeq
$~constant, and the pericentre distance goes as
\begin{equation}
p = a(1 - e) = 2a - a_0.
\end{equation}
The binary therefore coalesces once the original semimajor axis has
halved. We show below (equation~\ref{oneminuse}) that this occurs once
the secondary has 
absorbed the (negative) angular momentum of disc gas with mass
comparable to its own. If gas flows inwards through the circumbinary
disc at the rate $\dot M$ the timescale for coalescence is
\begin{equation}
t_{\rm co} \simeq {M_2\over \dot M}
\label{tco}
\end{equation}
We stress again that this does {\it not} require either hole to
accrete this gas but only to capture the gas into a bound orbit around
the hole. In particular, $\dot M$ can formally exceed the Eddington
accretion rate for either hole without affecting the orbital
shrinkage. If the accretion rates on to either hole were
super--Eddington then the captured gas would be blown away by
radiation pressure from the disc(s) around the secondary (and possibly
primary) black hole, without significantly changing the binary orbital
evolution (cf equation~\ref{oneminuse}).

The critical eccentricity separating cases where the binary remains
almost circular from those of growing eccentricity depends on the
surface density distribution of the circumbinary disc. In a real
three--dimensional disc with scale height $H$ and aspect ratio $H/R$
the surface density tails off over a length--scale $H(a) \sim (H/R)a$.
This suggests that the critical eccentricity dividing these two cases
is just
\begin{equation}
e_{\rm crit} \sim {H\over R}
\label{ecrit}  
\end{equation}
Thus any binary starting with $e > e_{\rm crit}$, or achieving it by
capturing a large mass (comparable to the secondary's) in one orbit,
must become very eccentric. We note that even a preceding episode of
prograde accretion can leave the binary with an eccentricity exceeding
$e_{\rm crit}$ (e.g. \citealt{Cuadraetal2009}), so that growth to 
high eccentricity is very likely if accretion is chaotic.

Our conclusions about eccentricity growth agree with those of
\citet{Dotti}, who considered a related but different problem.
A secondary black hole was injected with
significant eccentricity into a pre--existing dense circumnuclear disc
surrounding a primary black hole. The secondary's orbit
was initially retrograde with respect to this interior disc. However
the secondary was able to interact with enough gas in less than one
orbit that it cancelled all of its angular momentum. The secondary
then briefly had approximately zero angular momentum before capturing
more gas and so changing its angular momentum to prograde. In this
paper we restrict ourselves to accretion events on much smaller scales
than in \citet{Dotti}. This difference in lengthscales is important. At the
smaller scales we consider, a disc with mass $M_d \gg M_2$ would
probably be self--gravitating, and this change of angular momentum
sign is unlikely to occur. Our simulations agree with this conclusion:
once interaction with the disc has cancelled the orbital angular
momentum (and thus much of the orbital energy) of the secondary, the binary
coalesces.

\section{Orbital evolution with a retrograde circumbinary disc}
\label{orbevo}

We can now make analytic estimates of the orbital evolution as the
binary interacts with an exterior disc. For simplicity we again
assume that $q = M_2/M_1 \ll 1$, so that the secondary has specific
orbital energy and angular momentum $E, J^2$ where
\begin{equation}
E = -{GM_1\over 2a} = {1\over 2}v(r)^2 - {GM_1\over r}
\label{E}
\end{equation}
and 
\begin{equation}
J^2 = GM_1a(1-e^2)
\label{J}
\end{equation}

We have seen above that in many cases the binary eccentricity grows
quite strongly. In the limit the secondary interacts with the disc
only very near apocentre $r = a(1+e)$. Here its velocity $v_{\rm ap}$
is purely azimuthal, with
\begin{equation}
[a(1+e)v_{\rm ap}]^2 = J^2 
\end{equation}
which by equation~\ref{J} gives 
\begin{equation}
v_{\rm ap}^2 = {GM_1\over a}{1-e\over 1+e}
\label{vap}
\end{equation}
Near apocentre the secondary interacts with disc material moving with
azimuthal velocity $v_{\rm disc}< 0$, with
\begin{equation}
v_{\rm disc}^2 = {GM_1\over r}  = {GM_1\over a(1+e)}
\label{vdisc}
\end{equation}

We assume that a mass $\Delta M$ of disc matter is captured into orbit
about the secondary near apocentre, as we discussed above, so that all
of its orbital angular momentum is transferred to the secondary. To
allow for mass loss from the subsequent accretion process on to this
hole (if for example this is super--Eddington, or mass interacts
gravitationally with the secondary but is all accreted by the primary)
we assume that the effective mass of the hole plus disc becomes $M_2 +
\alpha\Delta M$, with $0 \leq \alpha \leq 1$. Then conservation of linear
momentum gives
\begin{equation}
M_2v_{\rm ap} - \Delta Mv_{\rm disc} = (M_2 + \alpha \Delta M)u
\label{mom}
\end{equation}
where $u$ is the new apocentre velocity of the secondary plus its
captured gas disc. The changes $\Delta E, \Delta a$ in orbital specific
energy and semi--major axis are given by
\begin{equation}
{GM_1\over 2a^2}\Delta a = \Delta E = {1\over2}u^2 - {1\over 2}v_{\rm ap}^2
\label{Deltas}
\end{equation}
Combining equations~\ref{vap},\ref{vdisc},\ref{mom} and \ref{Deltas} gives 
\begin{equation}
{\Delta a\over a^2} = {-2\over a(1+e)}{\Delta M\over M_2}[(1-e)^{1/2}
  + \alpha(1-e)]
\label{star}
\end{equation}
to lowest order in $\Delta M$. We noted above that if the interaction
is confined to the immediate vicinity of apocentre then the apocentre
distance $a_0 = a(1+e)$ stays constant in the subsequent
evolution. Then $1+e = a_0/a$ and $1-e = 2 - a_0/a$, so that the LHS
of equation~\ref{star} is simply $\Delta(1 - e)/a_0$ and therefore
equation~\ref{star} becomes 

\begin{equation}
\Delta (1-e) = -{2\Delta M\over M_2}[(1-e)^{1/2} + \alpha(1-e)].
\end{equation}

Using $M_2 = M_{20} + \alpha M$, where $M_{20}$ was the mass of the
secondary hole when $e$ was zero, and $M$ is the total mass since
transferred from the disc, this integrates to give
\begin{equation}
1-e = \biggl({M_{20}-M\over M_{20} + \alpha M}\biggr)^2,
\label{oneminuse}
\end{equation}
Hence in this approximation the binary coalesces (i.e. $1-e = 0$) once
the disc has transferred a mass equal to the secondary black hole,
i.e. after a time $t_{\rm co}$ (cf equation~\ref{tco}), {\it independently of the
fraction $\alpha$}. This means that mass loss has no effect in slowing
the inspiral. We note that equation~\ref{star} implies that the energy
dissipated in the shrinkage of an eccentric binary is 
\begin{equation}
-M_2\Delta E \simeq {GM_1\Delta M\over 2a}
\label{discdiss}
\end{equation}
per binary orbit, which is less than that produced by viscous
dissipation in the disc (this has to pull in a mass larger than
$\Delta M$ on each orbit).

All other cases give similar timescales $\sim t_{\rm co}$ for
coalescence. For example if the orbit stays circular we can set $e=0$
in equation~\ref{star} and find
\begin{equation}
{\Delta a\over a} = -2(1+\alpha){\Delta M \over M_2},
\end{equation}
so that $a \propto M_2^{-2(1+\alpha)}$. If the secondary gains the
transferred mass we have $\alpha = 1, a \propto M_2^{-4}$, while if
the transferred mass (but not the angular momentum) ends up on the
primary we have $\alpha = 0, a \propto M_2^{-2}$.  Shrinking the
binary from $a =1$~pc to $a = 10^{-2}$~pc, where gravitational
radiation rapidly coalesces it, requires the transfer of between 2 and
9 times the original mass of the secondary in the two cases. This is
larger than in the eccentric case because the torque on the binary now
decreases with $a$ (indeed the required mass would be formally
infinite without gravitational radiation).

\section{Simulations}
\label{Numerics}  

We now use Smoothed Particle Hydrodynamics (SPH) simulations to
compare with the analytical arguments above.
\subsection{Code setup}

We use a fully 3D conservative Lagrangian implementation of the SPH
algorithms, for example \citet{SandH2002}; \citet{Price2005}; \citet{Rosswog2009}. We
neglect gas self-gravity and so are able to use linked--lists of
particles rather than the usual tree for neighbour finding
\citep{DeeganPhD}. We assume an isothermal disc, i.e. $P =
c_{\mathrm{s}}^{2} \rho $, where the temperature and hence sound speed
$c_s$ is constant for all particles at all times. We make this choice
for simplicity, and anticipate that the interaction of gas with the
binary is not greatly affected by it.  In particular we noted in
Section~\ref{orbevo} above (cf equation~\ref{discdiss}) that binary shrinkage
does not 
greatly increase the heating of the disc.  We integrate a `live'
binary which feels the back reaction of the gas and generates
self--consistent orbits for the binary and the gas.

Our aim here is to understand the dynamical interaction of the disc
with the binary, rather than long--term behaviour such as the
timescales for coalescence, which we know depends on the total mass
transferred through the disc (see Section~\ref{orbevo} above). Accordingly we
use 
only the standard SPH artificial viscosity \citep{Rosswog2009} and do
not include a physical viscosity. Our simulations run only for a few
tens of binary orbits, so that the viscosity scheme employed should
not affect our results.

Our code units are dimensionless and take the initial separation and mass of
the binary as unity. The period of the binary in its initial configuration is
thus $2\pi$. The general disc setup is gas, of mass $M_{\rm d}$, spread from
$R_{\rm in}$ to $R_{\rm out}$ with a surface density following a power law in
radius, i.e. $\Sigma = \Sigma_{0}\left( R/R_{0} \right)^{-p}$ with $p$
typically $=1$. The initial vertical structure in the disc is a Gaussian. We
set the sound speed in the gas to a value ensuring that our neglect of
self--gravity is justified. Thus we arrange that the Toomre $Q$ parameter
\citep{Toomre} exceeds a minimum value ($> 1$) throughout the disc. Typically we
arrange $Q > 5$ for a disc of mass $M_{d}=10^{-2}$. The particles are all on
initially circular orbits.

The simulations detailed below were repeated with half a million,
one million and two million particles and were deemed to have
converged. All of the simulations detailed below initially contain
one million particles.\footnotemark[1]

\footnotetext[1]{Movies of these simulations are available and can be
    found at:\\
    \url{www.astro.le.ac.uk/users/cjn12/RetroBinaryMovies.html}}  

\subsection{Where does the mass go?}

Two main parameters govern the gas flow for a retrograde disc--binary
system: the binary mass ratio and the `capture' radius of each binary
component. As we discussed in \sref{retrobehaviour}, with mass ratios
$q \simeq 1$, both the primary and secondary can interact with the
inner edge of the disc so we should expect gas capture by both
objects. However if $q \ll 1$, an inward--moving disc reaches the
secondary before the primary. Then the secondary's effective radius
may be large enough for it to capture all of the gas. If instead this
radius is small the gas may be perturbed towards the primary rather
than captured. In Section~\ref{retrobehaviour} we estimated the critical value
dividing 
these cases (equation~\ref{rcrit}). We study the effects of changing
parameters here.

It is of course currently impossible to follow gas accretion down to
the innermost stable circular orbits of the two black holes while
simultaneously following the gas flow from the circumbinary disc. The
analytic work above suggests that the secondary hole is always
surrounded by a gas disc denser than the circumbinary disc and any
gas flow from it, which therefore captures any gas impacting it. If
initially there had been no gas around the secondary, gas entering
$R_{\rm capture} = aM_{2}/2M_{1}$ would be captured (but not
necessarily accreted). This gas would spread to form a disc around the
secondary, on a timescale short with respect to the mass transfer
timescale in the circumbinary disc. The secondary's disc then forms an
obstacle for further inflow from the circumbinary disc. As further
mass is captured in this way, this disc can become at most as large as
some fraction of the Roche lobe (cf equation~\ref{roche}). In the
language of planetary dynamics this condition is formally equivalent
to taking the secondary's Safronov number, defined as (planetary
escape velocity/orbital velocity)$^2$/2,
(e.g. \citealt{Safronovbook}; \citealt{HandB2007}) to be of order $2.5q^{2/3}$.

From this analysis we expect the capture radius of the holes to be of
order or smaller than the Roche lobe. For completeness we also
consider the possibility of still smaller capture radii. Once a
particle moves within the capture radius it is assumed to have
impacted upon the disc assumed to be present inside the capture
radius. This gas could be accreted by the black hole or expelled by
radiation pressure from the innermost parts of the disc. In either
case it is clear that once it has been `captured' it plays no
significant further role in the binary dynamics. Accordingly we add
its mass and momentum to the relevant hole and remove it from the
simulation.

\subsubsection{Secondary capture}
\label{sec} 

The first simulation has a mass ratio of $q=0.1$ and effective radii
0.15, 0.2 for the secondary and primary. Equation~\ref{rcrit} implies
that the secondary should capture the inner edge of the circumbinary
disc without significantly perturbing the remaining disc particles. From
equation~\ref{saf} the secondary's Safronov number is 2/3.

We start with the retrograde disc extending from 1 to 2 in radius, and a
circular binary.
\begin{figure}
  \begin{center}
  \includegraphics[angle=0,width=20pc]{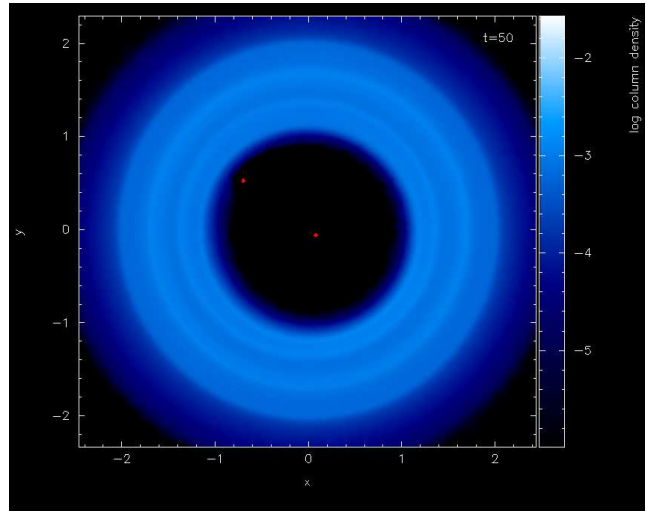}
  \caption{Image of the simulation from Section~\ref{sec} at time
    $t=50$. The binary is represented by the two dots. The axes
    are in code units with the log of the column density given by
    the bar.}
  \label{secaccn}
  \end{center}
\end{figure}
At first the disc inner edge is slightly perturbed, but not enough to
provide any noticeable accretion onto the primary. After one binary
orbit the system settles and the secondary smoothly captures the inner
edge of the disc. Figure~\ref{secaccn} shows the state of the
simulation at $t = 50$. At this time, of the particles accreted,
the secondary has captured $\sim 100\%$ (149759 particles) and the
primary has captured $\sim 0 \%$ (3 particles).

\subsubsection{Primary capture}
\label{pri} 

The second simulation also has a mass ratio of $q=0.1$, but this time
we use an effective radius of 0.05 for the secondary with again 0.2
for the primary capture radius, so that its Safronov number is $\Theta
= 2$.  Equation~(\ref{rcrit}) implies that the secondary should
perturb the disc particles significantly without directly accreting
all of them.

The initial disc is exactly the same as for the simulation in
Section~\ref{sec}. The binary is again initially circular.

\begin{figure}
  \begin{center}
  \includegraphics[angle=0,width=20pc]{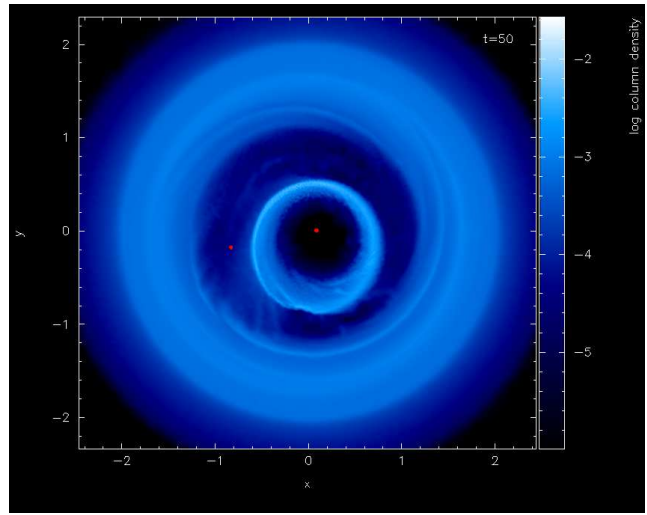}
  \caption{Image of the simulation from Section~\ref{pri} at time
    $t=50$. 
}
  \label{priaccn}
  \end{center}
\end{figure}

At first, the disc inner edge is significantly perturbed, with some
particles on orbits passing close to the primary. In the first 3 -- 4
orbits the gas flow is very chaotic. After this the system settles
into a quasi--steady state. Both holes capture from the retrograde
circumprimary disc (as shown in figure~\ref{priaccn}). The secondary
both disturbs and captures particles from the outer edge of the
circumprimary disc, and perturbs more particles into it. The
circumprimary disc is warped, eccentric and precessing.  This is
probably a result of particle noise destroying the symmetry about the
orbital plane, as might indeed happen in a realistic situation.

At time $t = 50$ the secondary has captured $\sim 95\%$ (147642) particles and
the primary has captured $\sim 5\%$ (8193 particles). So even in this case the
secondary still takes most of gas captured from the circumbinary disc. The
binary probably coalesces once the disc has transferred a mass $\sim M_2$, so
if the disc is more massive than this, the coalesced hole eventually accretes
the remainder. Clearly if the secondary's mass or its capture radius is made
arbitrarily small, it would accrete very little, and most of the mass would be
captured by the primary. However it is clear that this requires extreme
choices of these parameters.

\subsubsection{Dual Capture} 
\label{dual}

Here we look at the interaction of an near equal--mass binary with
the disc. We use $q=0.5$ and the same disc as in Sections~\ref{sec}
and \ref{pri}. Again the binary is initially circular. Here we use capture
radii of 0.01 for both holes to allow formation of circumprimary and
circumsecondary discs. 

\begin{figure}
  \begin{center}
  \includegraphics[angle=0,width=20pc]{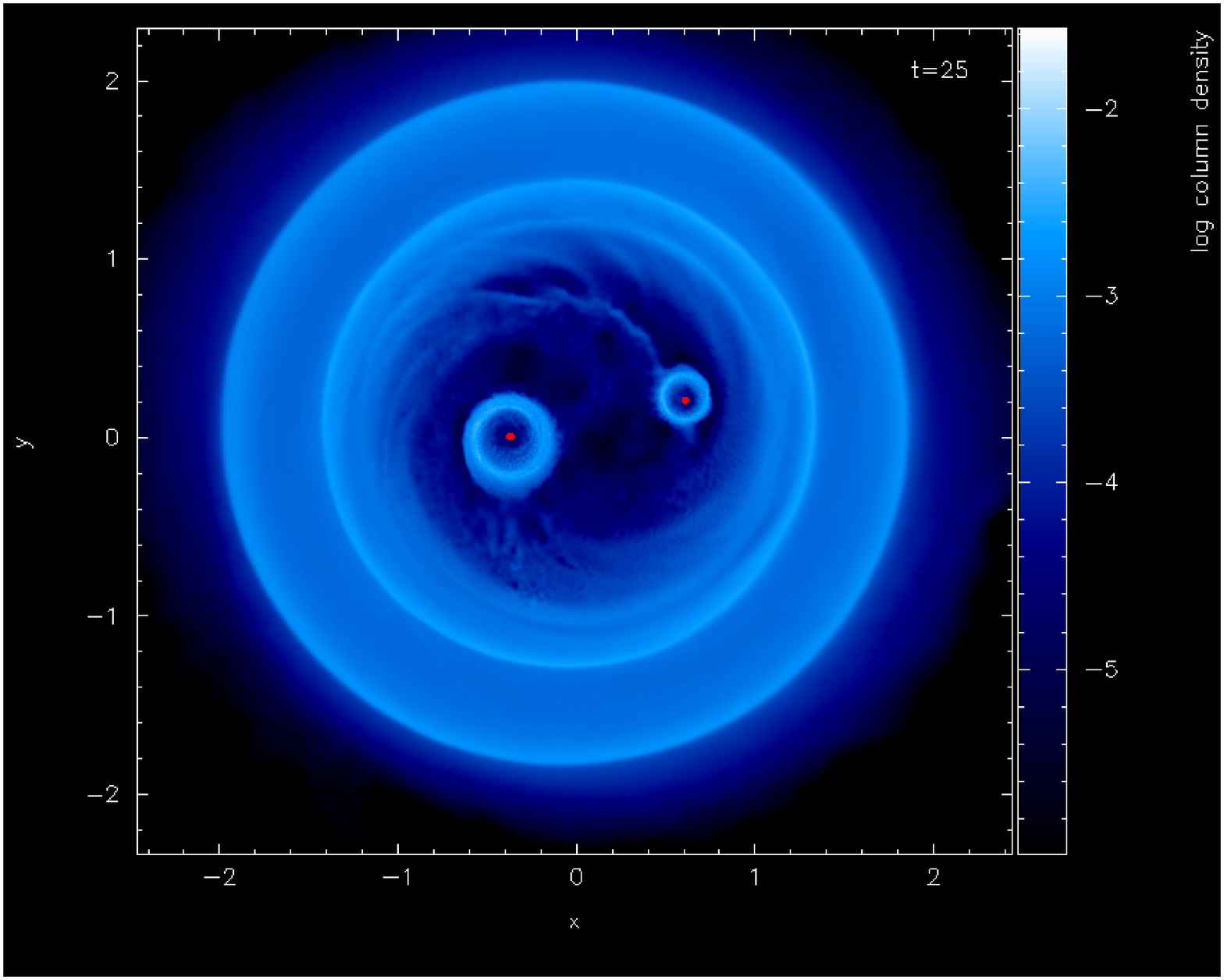}
  \includegraphics[angle=0,width=20pc]{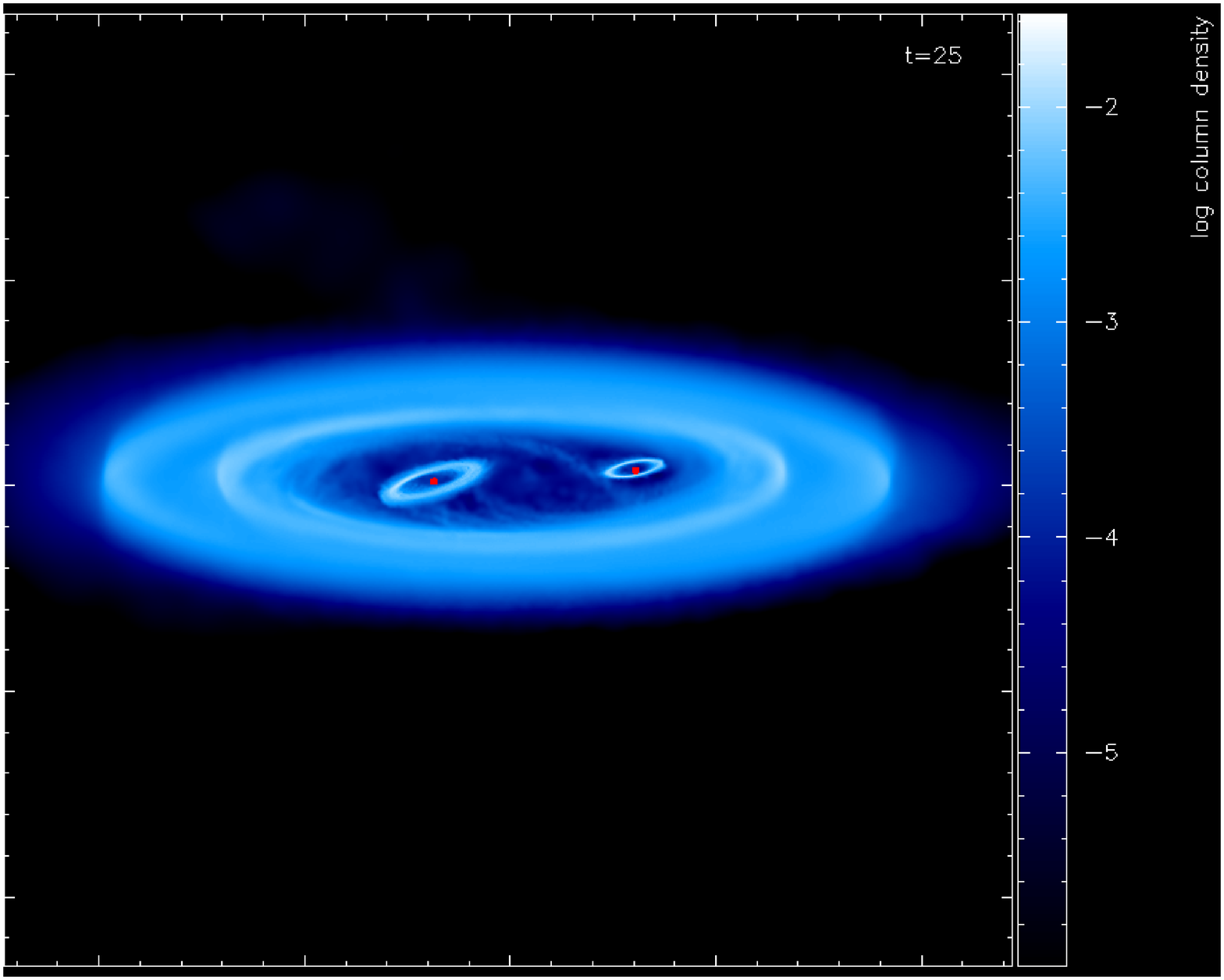}
  \caption{Images of the simulation from Section~\ref{dual} at time
    $t=25$. 
The upper panel shows the disc
    face on. The lower panel shows the same plot but viewed at an
    angle of $15^{\circ}$ to the plane of the disc.}
  \label{both}
  \end{center}
\end{figure}

During the initial 3 -- 4 binary orbits the flow is very chaotic, with
mass captured by both holes. As more gas is captured (but this time
not `accreted') by both holes, circumprimary and circumsecondary discs
are formed which persist throughout the simulation. The discs are
supplied by streams from the circumbinary disc. We show the simulation
at time $t=25$ in figure~ref{both}. At this time the primary has
captured $48\%$ (57969) and the secondary $52\%$ (62189) of the
accreted particles. Together with the previous simulation, this shows
that the primary can gain significant mass only if the mass ratio is
close to unity. We note again that the discs are not planar, and show
a significant tilt with respect to the binary plane. Again this is
probably a result of particle noise removing the symmetry about the
orbital axis. Globally angular momentum is conserved but locally the
streams that supply these discs need not be planar.

\subsection{Eccentricity growth}
\label{Eccngd}

The analytic arguments of Section~\ref{sec4} suggest that capture from a
retrograde circumbinary disc at apocentre and pericentre of the binary
orbit increase and decrease its eccentricity respectively. Here we
show three SPH experiments exploring this.

The first simulation has an initially circular binary, with
$q=10^{-3}$, where the secondary is embedded just inside the inner
edge of a retrograde disc. In code units the disc is spread radially
from 0.8 to 1.5. The disc has an initial mass $M_{d} = 10^{-2}$.

The second simulation also starts with the secondary embedded in the
disc. However this time the inner edge of the disc extends much
further in, to 0.1.

The third simulation starts with a retrograde disc interior to the
binary, i.e. a circumprimary disc extending from 0.1 to 1.0. In this
case we start with initial eccentricity 0.5 and the binary at
apocentre -- so the secondary begins outside the circumprimary disc
but plunges into it before it reaches pericentre.

All of these simulations have capture radii 0.1, 0.01 for the
primary and secondary respectively.

\begin{figure}
  \begin{center}
  \includegraphics[angle=0,width=20pc]{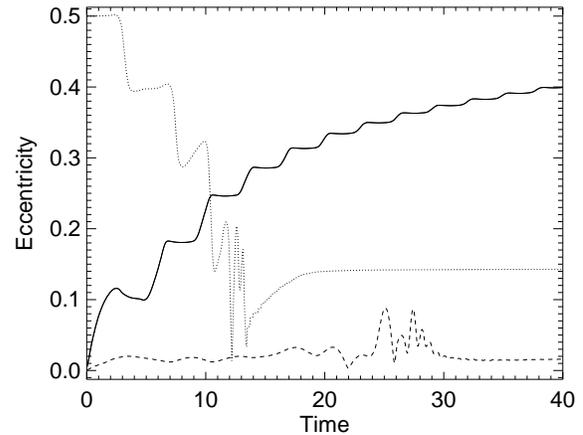}
  \caption{Eccentricity growth and decay of the simulations in
    Section~\ref{Eccngd}. Time is in code units. The solid line is the
    first simulation (capture at apocentre but not at pericentre), the
    dashed line is the second simulation (capture at both pericentre and 
    apocentre) and the dotted line is the third simulation (capture at
    pericentre and not apocentre) }
  \label{eccngrowdecay}
  \end{center}
\end{figure}

We show in figure~\ref{eccngrowdecay} how the eccentricity of the
binary evolves in the three cases. In the first simulation the binary
captures enough gas at apocentre on the first orbit to plunge inside
the inner edge of the disc. This means that the eccentricity growth at
apocentre cannot be moderated by decay at pericentre. The eccentricity
therefore grows as shown in figure~\ref{eccngrowdecay}. In the second
simulation the secondary is still capturing when it reaches pericentre
and so the eccentricity decays. It is notable that the eccentricity never
returns precisely back to zero in this case. This happens 
because the secondary captures unequal amounts of gas at apo-- and
pericentre. This is reasonable, as its velocity is higher at
pericentre and so it has less time to interact with the gas, In
addition the mass and angular momentum at each radius is not constant.

In the third simulation the secondary can initially only capture from
the circumprimary disc at pericentre and so the eccentricity
decays. After approximately $3 - 4$ orbits the secondary has damped
out most of its initial eccentricity to an almost circular orbit. After a
short interaction with the inner parts of the disc it reaches
its inner edge. Here there is a brief interval of eccentricity growth
as it captures from this inner edge of a (now) circumbinary disc.

All of the results of these simulations agree with the analytic
arguments of Section~\ref{sec4}.

\section{Discussion}

We have seen that accretion from a retrograde circumbinary disc can be
considerably more effective in shrinking a supermassive black hole
binary system than accretion from a corresponding prograde
disc. Coalescence occurs if the secondary black hole captures angular
momentum from a gas mass comparable to its own.

The reason for this effectiveness is the absence of orbital resonances
in a retrograde disc. This cannot react to tidal torques from the
binary and become a decretion disc, which is what tends to slow the
evolution in the prograde case
(\citealt{Lodatoetal2009}; \citealt{Cuadraetal2009}; \citealt{Dottietal2007}; \citealt{MandM2008}).
Instead, the disc directly feeds negative orbital angular momentum to
the secondary black hole. An important aspect here is that there is no
restriction on the rate at which this can occur. In particular gas can
flow inwards at rates higher than Eddington for even the
primary black hole without any significant effect on the amount of gas
required to coalesce the binary.

Capture to the primary hole is negligible unless the mass ratio is
very close to unity, and is never dominant (even if `capture' actually
leads to `accretion', which is not required). This may be important in
interpreting attempts to observe merger events. In many cases the
binary may develop very high eccentricity before gravitational wave
emission coalesces it. The latter rapidly damps the eccentricity
during the final inspiral, but leaves a residual value which is likely
to be significant in any merger event detectable by gravitational wave
observatories such as LISA. If the secondary hole actually accretes
the captured mass it will acquire a large spin (Kerr parameter $a
\simeq 1$) antiparallel to its binary orbit, which would be potentially
detectable in the LISA waveform. Moreover if the mass ratio is very
close to unity it is conceivable that the primary hole might do the
same. This could in principle favour high black hole spin as a result
of major mergers (e.g. in giant ellipticals), as sometimes
proposed. However we should recall that the holes may well {\it not}
gain much mass, particularly if this is captured at super--Eddington
rates, and the primary definitely does not gain mass (and thus spin
up) unless the mass ratio is very close to unity. For these reasons it
seems unlikely that coalescences induced by retrograde gas flows
produce rapidly spinning merged holes except in rare cases.

We caution finally that so far we have not considered two important
effects.  First, we assumed that the retrograde disc was coplanar with
the binary, whereas in reality counter--alignment must occur over a
viscous dissipation timescale. Second, self--gravity can deplete the
circumbinary disc of gas and reduce its ability to shrink the binary
(cf. \citealt{Lodatoetal2009} in the prograde case). Since coalescence
by a retrograde disc requires $M_d \gtrsim M_2$, and self--gravity
effects appear unless $M_d \lesssim (H/R)(M_1 + M_2)$, where $H/R$ is
the disc aspect ratio, it appears that the last parsec problem is so
far alleviated only for mergers with $M_2/M_1 \lesssim H/R$. This
paper does suggest how coalescence might work for larger mass
ratios. The secondary hole has to absorb negative angular momentum
from a gas mass comparable to its own, preferably in an eccentric
orbit. Although gas self--gravity is inevitably important in such an
event, a retrograde flow has the advantage that there is no limit on
the mass inflow rate. We shall investigate these two effects in future
work.

\section*{Acknowledgments}
\label{acknowledgements}
We thank Walter Dehnen and Frazer Pearce for useful discussions on
SPH.  We thank the referee for encouraging us to investigate the conditions
for orbital eccentricity growth.
We acknowledge the use of \textsc{splash} \citep{Price2007} for
the rendering of the SPH plots. CJN and PJC acknowledge STFC
studentships. Research in theoretical astrophysics at Leicester is
supported by an STFC Rolling Grant.
\bibliographystyle{mn2e} 
\bibliography{nixon}

\begin{thebibliography}{}

\bibitem[\protect\citeauthoryear{{Armitage} \& {Natarajan}}{{Armitage} \&
  {Natarajan}}{2005}]{ArmNar2005}
{Armitage} P.~J.,  {Natarajan} P.,  2005, ApJ, 634, 921

\bibitem[\protect\citeauthoryear{{Berti} \& {Volonteri}}{{Berti} \&
  {Volonteri}}{2008}]{Bertvol}
{Berti} E.,  {Volonteri} M.,  2008, ApJ, 684, 822

\bibitem[\protect\citeauthoryear{{Cuadra}, {Armitage}, {Alexander} \&
  {Begelman}}{{Cuadra} et~al.}{2009}]{Cuadraetal2009}
{Cuadra} J.,  {Armitage} P.~J.,  {Alexander} R.~D.,    {Begelman} M.~C.,  2009,
  MNRAS, 393, 1423

\bibitem[\protect\citeauthoryear{{Deegan}}{{Deegan}}{2009}]{DeeganPhD}
{Deegan} P.,  2009, PhD thesis, Univ.~Leicester, 2009

\bibitem[\protect\citeauthoryear{{Dotti}, {Colpi}, {Haardt} \& {Mayer}}{{Dotti}
  et~al.}{2007}]{Dottietal2007}
{Dotti} M.,  {Colpi} M.,  {Haardt} F.,    {Mayer} L.,  2007, MNRAS, 379, 956

\bibitem[\protect\citeauthoryear{{Dotti}, {Ruszkowski}, {Paredi}, {Colpi},
  {Volonteri} \& {Haardt}}{{Dotti} et~al.}{2009}]{Dotti}
{Dotti} M.,  {Ruszkowski} M.,  {Paredi} L.,  {Colpi} M.,  {Volonteri} M.,
  {Haardt} F.,  2009, MNRAS, 396, 1640

\bibitem[\protect\citeauthoryear{{Escala}, {Larson}, {Coppi} \&
  {Mardones}}{{Escala} et~al.}{2005}]{Escala2005}
{Escala} A.,  {Larson} R.~B.,  {Coppi} P.~S.,    {Mardones} D.,  2005, ApJ,
  630, 152

\bibitem[\protect\citeauthoryear{{Ferrarese} \& {Merritt}}{{Ferrarese} \&
  {Merritt}}{2000}]{FandM2000}
{Ferrarese} L.,  {Merritt} D.,  2000, ApJL, 539, L9

\bibitem[\protect\citeauthoryear{{Gebhardt}, {Bender}, {Bower}, {Dressler},
  {Faber}, {Filippenko}, {Green}, {Grillmair}, {Ho}, {Kormendy}, {Lauer},
  {Magorrian}, {Pinkney}, {Richstone} \& {Tremaine}}{{Gebhardt}
  et~al.}{2000}]{Gebhardtetal2000}
{Gebhardt} K.,  {Bender} R.,  {Bower} G.,  {Dressler} A.,  {Faber} S.~M.,
  {Filippenko} A.~V.,  {Green} R.,  {Grillmair} C.,  {Ho} L.~C.,  {Kormendy}
  J.,  {Lauer} T.~R.,  {Magorrian} J.,  {Pinkney} J.,  {Richstone} D.,
  {Tremaine} S.,  2000, ApJL, 539, L13

\bibitem[\protect\citeauthoryear{{Hansen} \& {Barman}}{{Hansen} \&
  {Barman}}{2007}]{HandB2007}
{Hansen} B.~M.~S.,  {Barman} T.,  2007, ApJ, 671, 861

\bibitem[\protect\citeauthoryear{{H{\"a}ring} \& {Rix}}{{H{\"a}ring} \&
  {Rix}}{2004}]{HandR2004}
{H{\"a}ring} N.,  {Rix} H.,  2004, ApJl, 604, L89

\bibitem[\protect\citeauthoryear{{Hobbs}, {Nayakshin}, {Power} \&
  {King}}{{Hobbs} et~al.}{2010}]{Hobbs}
{Hobbs} A.,  {Nayakshin} S.,  {Power} C.,    {King} A.,  2010, ArXiv e-prints

\bibitem[\protect\citeauthoryear{{King} \& {Pringle}}{{King} \&
  {Pringle}}{2006}]{KandP2006}
{King} A.~R.,  {Pringle} J.~E.,  2006, MNRAS, 373, L90

\bibitem[\protect\citeauthoryear{{King}, {Pringle} \& {Hofmann}}{{King}
  et~al.}{2008}]{Kingetal2008}
{King} A.~R.,  {Pringle} J.~E.,    {Hofmann} J.~A.,  2008, MNRAS, 385, 1621

\bibitem[\protect\citeauthoryear{{Lodato}, {Nayakshin}, {King} \&
  {Pringle}}{{Lodato} et~al.}{2009}]{Lodatoetal2009}
{Lodato} G.,  {Nayakshin} S.,  {King} A.~R.,    {Pringle} J.~E.,  2009, MNRAS,
  398, 1392

\bibitem[\protect\citeauthoryear{{MacFadyen} \&
  {Milosavljevi{\'c}}}{{MacFadyen} \& {Milosavljevi{\'c}}}{2008}]{MandM2008}
{MacFadyen} A.~I.,  {Milosavljevi{\'c}} M.,  2008, ApJ, 672, 83

\bibitem[\protect\citeauthoryear{{Mayer}, {Kazantzidis}, {Madau}, {Colpi},
  {Quinn} \& {Wadsley}}{{Mayer} et~al.}{2007}]{Mayer2007}
{Mayer} L.,  {Kazantzidis} S.,  {Madau} P.,  {Colpi} M.,  {Quinn} T.,
  {Wadsley} J.,  2007, Science, 316, 1874

\bibitem[\protect\citeauthoryear{Milosavljevic \& Merritt}{Milosavljevic \&
  Merritt}{2003}]{MandM2003}
Milosavljevic M.,  Merritt D.,  eds, 2003 American Institute of Physics: The
  Astrophysics of Gravitational Wave Sources: AIP Conference Proceedings

\bibitem[\protect\citeauthoryear{{Papaloizou} \& {Pringle}}{{Papaloizou} \&
  {Pringle}}{1977}]{PandP1977}
{Papaloizou} J.,  {Pringle} J.~E.,  1977, MNRAS, 181, 441

\bibitem[\protect\citeauthoryear{{Price}}{{Price}}{2005}]{Price2005}
{Price} D.,  2005, arXiv:astro-ph/0507472

\bibitem[\protect\citeauthoryear{{Price}}{{Price}}{2007}]{Price2007}
{Price} D.~J.,  2007, Publications of the Astronomical Society of Australia,
  24, 159

\bibitem[\protect\citeauthoryear{{Rosswog}}{{Rosswog}}{2009}]{Rosswog2009}
{Rosswog} S.,  2009, New Astronomy Review, 53, 78

\bibitem[\protect\citeauthoryear{{Safronov}}{{Safronov}}{1972}]{Safronovbook}
{Safronov} V.~S.,  1972, {Evolution of the protoplanetary cloud and formation
  of the earth and planets.}

\bibitem[\protect\citeauthoryear{{Springel} \& {Hernquist}}{{Springel} \&
  {Hernquist}}{2002}]{SandH2002}
{Springel} V.,  {Hernquist} L.,  2002, MNRAS, 333, 649

\bibitem[\protect\citeauthoryear{{Toomre}}{{Toomre}}{1964}]{Toomre}
{Toomre} A.,  1964, ApJ, 139, 1217

\end{thebibliography}

\appendix
\section{Binary--Disc interaction}
\label{appA}

To understand the interaction between the gas and the binary we
consider the case where the disc particles can be perturbed before
impacting upon the secondary. The relative velocity of a disc
particle before interacting is $2V$. After the interaction it gains
a radial velocity (\eref{Ur}). In the frame of $M_{2}$ the energy
of the particle is conserved and so
\begin{equation}
  \left(2V \right)^{2} = U_{R}^{2} + U_{T}^2.
\end{equation}  
so that 
\begin{equation}
  U_{T} \approx 2V\left[1-\frac{G^{2}M_{2}^2}{b^2V^4} \right].
\end{equation}  
Back in the inertial frame, this implies that after the interaction
the radial velocity of the disc particle is $U_{R}$ and the azimuthal
velocity is $U_{\phi} = U_{T}-U$, i.e.
\begin{equation}
  U_{\phi} = V-\frac{G^{2}M_{2}^{2}}{b^2V^4}.
\end{equation}  
The particle was on an orbit with eccentricity $e=0$, specific energy
$E=GM_{1}/a$ and specific angular momentum $h=\left(GM_{2}R
\right)^{1/2}$. After the interaction the particle's specific kinetic
energy is $T'$ where
\begin{equation}
  T' = \frac{U_{\phi}^2}{2} + \frac{U_{R}^2}{2} = \frac{V^{2}}{2} +
  \frac{2G^{2}M_{2}^{2}}{b^{2}V^{2}}.
\end{equation}
Thus after the interaction the specific energy of the particle orbit
is increased to
\begin{equation}
  E' = E + \Delta E,
\end{equation}
where
\begin{equation}
  \Delta E = \frac{G^{2}M_{2}^{2}}{b^{2}V^{2}}
\end{equation}
and the particle's specific angular momentum is decreased to 
\begin{equation}
  h' = h - \Delta h,
\end{equation}
where 
\begin{equation}
  \Delta h = a\frac{G^{2}M_{2}^{2}}{b^{2}V^{3}}.
\end{equation}

We note that the angular momentum of the particle has the opposite
sign to that of the secondary, as it must.

The perturbed particle now has semi-major axis 
larger than $a$, by an amount $\Delta a = a\left(\Delta E / E
\right)$. Thus 
\begin{equation}
  \Delta a = a\frac{G^{2}M_{2}^{2}}{b^{2}V^{4}},
\end{equation}
and also for comparison 
\begin{equation}
  \frac{\Delta a}{b} = \frac{a^3}{b^3} \frac{M_{2}^{2}}{M_{1}^{2}}.
\end{equation}
The eccentricity of the particle's new orbit is given by 
\begin{equation}
  e \approx \frac{\Delta h}{h} - \frac{\Delta E}{2E} = \frac{\Delta h}{h} +
  \frac{\Delta a}{2a}.
\end{equation}
So we find
\begin{equation}
  e \approx 3\frac{G^{2}M_{2}^{2}}{b^{2}V^{4}}.
\end{equation}

From this we conclude that the interaction with the binary increases
the particle's orbital energy and decreases its orbital angular
momentum, leading to an eccentric orbit. Now there are two
possibilities for the disc. Either the disc now becomes eccentric
itself. Or the orbits of the perturbed particles begin to cross and
therefore the particles collide, lose energy and share angular
momentum.

As the disc can radiate away the excess energy associated with the
eccentric disc orbits (cf equation~\ref{discdiss}) it is clear that the overall effect on the disc
of the interaction with the binary is to shrink the disc.  Similarly
for the binary we conclude that the binary loses both orbital energy
and angular momentum and hence shrinks.

\end{document}